\documentclass{mem}
\usepackage{natbib}
\usepackage{txfonts}
\usepackage{balance}
\usepackage{graphicx}
\usepackage[a4paper]{hyperref}
\begin{document}
\def\mbh{M_\mathrm{BH}}
\def\rg{R_\mathrm{g}}
\def\teff{$T\rm_{eff }$}
\def\kms{$\mathrm {km s}^{-1}$}
\def\rd{\mathrm{d}}
\def\bs{\!\!}
\def\mcnd{M_\mathrm{d}}
\def\rcnd{R_\mathrm{d}}
\def\msun{M_\odot}
\def\sgra{SgrA$^\star$}

\title{SMBH Feeding and Star Formation in Massive Accretion Discs}

\subtitle{}

\author{L. \v{S}ubr\inst{1,2}}
%\offprints{L. \v{S}ubr}

\institute{
Faculty of Mathematics and Physics, Charles University,
V Hole\v{s}ovi\v{c}k\'ach 2, CZ-18000, Praha, Czech Republic
\and
Astronomical Institute, Academy of Sciences,
Bo\v{c}n\'{\i}~II, CZ-14131~Praha, Czech Republic
\email{subr@sirrah.troja.mff.cuni.cz}
}

\authorrunning{\v{S}ubr}

\titlerunning{SMBH Feeding and Star Formation}

\abstract{
Galactic nuclei are unique laboratories for the study of processes connected
with the accretion of gas onto supermassive black holes. At the same time,
they represent challenging environments from the point of view of stellar
dynamics due to their extreme densities and masses involved. There is a growing
evidence about the importance of the mutual interaction of stars with gas in
galactic nuclei. Gas rich environment may lead to stellar formation which,
on the other hand, may regulate accretion onto the central mass. Gas in 
the form of massive torus or accretion disc further influences stellar
dynamics in the central parsec either via gravitational or hydrodynamical 
interaction. Eccentricity oscillations on one hand and energy dissipation on 
the other hand lead to increased rate of infall of stars into the supermassive
black hole. Last, but not least, processes related to the stellar dynamics may
be detectable with forthcoming gravitational waves detectors.

}
\maketitle{}

\section{Introduction}
Standard model of active galactic nuclei consists of a supermassive black hole
(SMBH) of mass $\mbh$ surrounded by an accretion disc which extends to
$\sim 10^4 - 10^5\rg;\;\rg\equiv G\mbh/c^2$. Further away, at
$r\gtrsim10^6 \rg$, there is assumed
to be an obscuring massive molecular torus. Within the radius of
$\sim 10^6 - 10^7\rg$, there is a dense stellar cusp, of the total mass
comparable to $\mbh$. In spite of its extreme density which may exceed $10^8$
stars per cubic parsec in the centre, the two-body
relaxation time in this region is of the order of Gyrs. Hence, on the orbital
(crossing) time scale, the stars follow nearly Keplerian orbits in the
gravitational field of the SMBH. Interaction of stars with the axisymmetric
gaseous structures leads to a secular orbit evolution. Its characteristic time
may be well below the relaxation time and, therefore, it can lead to observable
effects.

In this contribution we briefly introduce various modes of the mutual
interaction of the stars and gas in galactic nuclei and discuss possible
consequences. In Section~\ref{sec:gc} we present our results within the context
of the nucleus of the Milky Way.

\section{Gravitational interaction}
\label{sec:gravity}
Gravitational field of the SMBH as well as the mean field of the star cluster
are assumed to be spherically symmetric, which implies conservation of the
vector of the
angular momentum. This is no longer true once a non-spherical perturbation to
the gravitational field is introduced. We will concentrate on the case of
an additional component of the gravitational field due to the accretion disc
or molecular torus for which we assume an axial symmetry. Then, one component
of the angular momentum vector, $L_z$, parallel to the symmetry axis will be
conserved, but the eccentricity may change.
%Numerical integrations show that for
%a large set of initial conditions, beside the energy and $L_z$ there exists
%a third integral of motion which poses an additional constraint on the secular
%evolution of the star's orbital elements.

This process, sometimes referred to as Kozai oscillations, is well described by
means of the perturbation Hamiltonian theory for an analogical (reduced)
hierarchical triple system \citep{kozai62,lidov62}. For the simplest case of
perturbation potential due to an infinitesimally narrow ring of mass $\mcnd$
and radius $\rcnd$, the equations of motion in the first approximation read:
\begin{eqnarray}
T_\mathrm{K}\,\eta\,\,\frac{\rd e}{\rd t} &\bs=\bs&
 {\frac{15}{8}}\,e\,\eta^2\,\sin2\omega\,\sin^{2}i\,,
 \label{eq:dedt} \\
T_\mathrm{K}\,\eta\,\,\frac{\rd i}{\rd t} &\bs=\bs&
 -\frac{15}{8}\,e^2\,\sin2\omega\,\sin i\,\cos i\,,
 \label{eq:didt} \\
T_\mathrm{K}\,\eta\,\,\frac{\rd\omega}{\rd t} &\bs=\bs&
 \frac{3}{4}\left\{ 2\eta^2+5\sin^{2}\omega\left[e^{2}-\sin^{2}i\right]\right\} ,
% + T_\mathrm{K} \frac{\rg}{a\,\eta}\,\frac{6\pi}{P}\,.
\label{eq:dodt} \\
T_\mathrm{K}\,\eta\,\,\frac{\rd\Omega}{\rd t} &\bs=\bs&
 -\frac{3}{4}\cos i \left[1+4e^2-5e^2\cos^{2}\omega\right] ,
\label{eq:dOdt}
\end{eqnarray}
where $a,\,e,\,i,\,\omega$ and $\Omega$ are the mean orbital semi-major
axis, eccentricity, inclination with respect to the plane of symmetry of the
perturbing potential, argument of the pericentre and the longitude of the
ascending node, respectively. We denote $\eta\equiv\sqrt{1-e^2}$ and
\begin{equation}
T_\mathrm{K} \equiv\frac{\mbh}{\mcnd} \, \frac{\rcnd^3}{a\sqrt{G\mbh a}}\;,
\label{eq:TK}
\end{equation}
which is the characteristic time of the evolution of the orbital elements.
Equations (\ref{eq:dedt}) -- (\ref{eq:dodt}) have two integrals of motion,
\begin{equation}
 C_1\!=\!\eta\sin i\; \mbox{ and }\;
 C_2\!=\!(5\sin^2 i \sin^2 \omega - 2)e^2,
\end{equation}
which implies that three orbital elements, $e,\,i$ and $\omega$, change
periodically with equal period and phase. For low values of $C_1$, the
amplitude of the oscillations of eccentricity may, in terms of $\eta$, reach
several orders of magnitude. Therefore, we expect enhanced rate of events
connected with extreme eccentricities, e.g. tidal disruptions of stars.

In the case of the galactic nuclei, there are, however, two additional
perturbations to the dynamics of individual stars, that tend to diminish
the amplitude of the eccentricity oscillations. It is the mean potential
of the stellar cusp and the relativistic pericentre advance. For some sets
of parameters, both of them lead to substantial increase of
$|\rd\omega / \rd t|$, i.e. to the shortening of the period of the oscillations
and also to damping of their amplitude \citep{blaes02,ivanov05}.

\begin{figure}[t]
\includegraphics[width=\columnwidth]{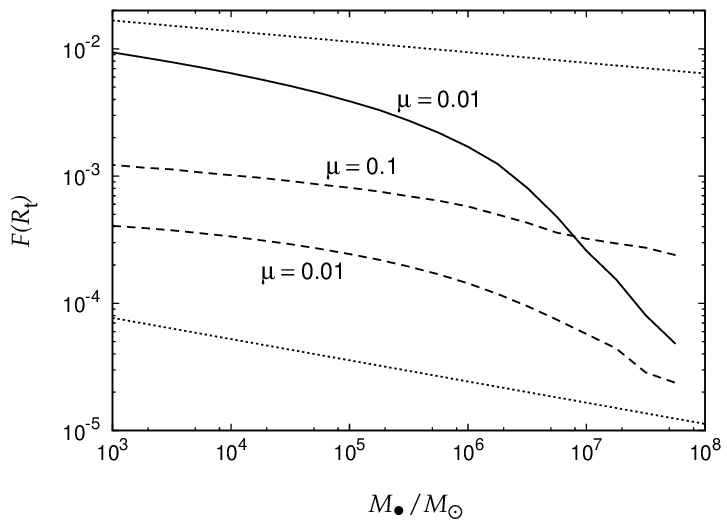}
\caption{\footnotesize Size of the loss cone of tidal disruptions as a fraction
of the phase space. Lower dotted line represents size of the loss cone for an
unperturbed case. Upper dotted line corresponds to an estimate for the case
of perturbation due to the ring. Model represented by the solid
line contains the 1.PN correction in addition. Dashed lines stand
for complex models, including the gravity of the stellar cusp.}
\label{fig:F_Rt}
\end{figure}
In \citet{ks07} we have compared the volume of the loss cone of the central
star cluster occupied by stars that reach the SMBH within the tidal radius,
$R_\mathrm{t}$. We have
considered a \citet{bahcall76} profile of the cusp and various disturbing
components to the central Keplerian potential, i.e. the axially symmetric
ring or disc, first post-Newtonian correction to the central potential and
the spherically symmetric mean potential of the stellar cusp. The results are
presented in Fig.~\ref{fig:F_Rt}. When considering only the ring as a
perturbation, the loss cone is more than two orders of magnitude larger in
comparison to the unperturbed case. Relativistic pericentre advance damps the
Kozai oscillations for $a<a_\mathrm{t}$,
\begin{equation}
 a_\mathrm{t}^7\approx{\textstyle{\frac{32}{9}}}\,\rcnd^6\,\rg^2\,
 R_\mathrm{t}^{-1}\,\mu^{-2}\,,
\label{eq:a_term}
\end{equation}
where we denote $\mu\equiv \mcnd/\mbh$. This transforms into the break
of the solid line in Fig.~\ref{fig:F_Rt} at $\mbh\approx10^6\msun$.
The pericentre shift caused by the stellar cusp, mass of which was set
equal to $\mbh$ in Fig.~\ref{fig:F_Rt}, leads to a
systematic decrease of the loss cone over the whole range of the black hole
mass. We may conclude that axisymmetric perturbation to the potential of
the SMBH may increase the number of the tidally disrupted stars by a factor
$\gtrsim10$. This number may by further increased if the source of the
axisymmetric component of the potential extends over a large range of radii
\citep{ks07}.

\section{Hydrodynamical interaction}
The dynamical processes described in the previous Section do not (in the first
approximation) change energy of individual stars. On the other hand, repetitive
passages of the stars through the accretion disc lead to dissipation of their
kinetic energy \citep{syer91,vk98,sk99}. Being highly supersonic, the
dissipational efficiency of the star--disc collisions is assumed to be
proportional to the physical cross-section of the star and the column density
of the accretion flow. For a standard gas pressure dominated \citet{ss73}
accretion disc the characteristic time of the orbital decay can be estimated
as \citep{skh04}:
\begin{equation}
 t_\mathrm{coll} \approx M_8 \left( \frac{\Sigma_\ast}{\Sigma_\odot} \right)
 \left( \frac{a}{\rg} \right)^{9/4}\mathrm{yr}\,,
\end{equation}
where $\Sigma_\ast \equiv M_\ast / R_\ast^2$ is the mean column density of the
star and $M_8 \equiv \mbh/10^8 \msun$. Beside the shrinking of the semi-major
axis this interaction leads to circularisation of the orbit and its decay into
the plane of the disc \citep{ks01}. We distinguish two different modes of
subsequent migration towards the centre once the star gets embedded in the
disc. If the star's Roche radius is larger than
the half-thickness of the disc, it opens a permanent gap in it and follows its
slow radial inflow. If the gap is not opened, the star excites density waves
in the gas which leads to larger transfer of angular momentum and, consequently,
faster inward migration \citep{lin86,ward86} on the time-scale of
\begin{equation}
 t_\mathrm{drag} \approx 100\, M_8^{1/10}
 \left( \frac{M_\ast}{M_\odot} \right)^{-1}
 \left( \frac{a}{\rg} \right)^{1/2}\!\mathrm{yr}\,.
\end{equation}

The inner region of the star cluster interacting with the accretion disc is
assumed to be continuously supplied with new stars on the relaxation
time-scale:
\begin{equation}
 t_\mathrm{relax} \approx 10^8\, n_6^{-1} M_8^{7/8}
 \left( \frac{M_\ast}{M_\odot} \right)^{-2}
 \left( \frac{r}{\rg} \right)^{1/4}\!\mathrm{yr}\,,
\end{equation}
where $n_6 \equiv n_0 / (10^6 \mathrm{pc}^{-2})$ is the stellar density.
Condition $t_\mathrm{coll} < t_\mathrm{relax}$ defines radius $R_\mathrm{out}$
of the region which
will be emptied due to the drag imposed by the disc upon the stars. The
inflow rate of stars towards the centre through the boundary of this region
is \citep{skh04}
\begin{equation}
 \dot{N} \approx 10^{-2}\, n_6^2\, M_8^{5/4}
 \left( \frac{M_\ast}{M_\odot} \right)^{2}
 \frac{R_\mathrm{out}}{10^4\rg}\, \mathrm{yr}^{-1}\,.
\end{equation}

\section{The Galactic Centre}
\label{sec:gc}
The closest galactic nucleus is that of the Milky Way. It harbours a SMBH
of mass $\mbh\approx 3.5\times 10^6\msun$ \citep{genzel03,ghez03} which
manifests itself as a radio source \sgra. In spite of that current activity
of \sgra is highly sub-Eddington, there is an indirect evidence for an
existence of massive accretion disc few million years ago: The near infrared
observations of the central parsec have revealed a numerous population of
young stars. Moreover, it has been found \citep{levin03} that considerable fraction
of them forms a relatively thin disc rotating coherently around the SMBH.
Fragmentation of a self-gravitating gaseous disc of mass $\gtrsim10^4\msun$
is currently one of the most popular explanation of the origin of this
young stellar structure.

Stellar disc induces axisymmetric perturbation to the potential of the SMBH.
Therefore, we assume that it leads to the eccentricity oscillations of
late-type stars from the embedding spherical stellar cusp. In \citet{ks07} we
have estimated that up to 100 stars (i.e. $\sim$ 2\% of the stars from the
region with $r\lesssim 0.3\mathrm{pc}$ may have been tidally disrupted due to
this process within the lifetime of the disc. Another source of axisymmetric
perturbation to the gravity of the central black hole is a circum-nuclear
disc (CND), a molecular torus of mass $\gtrsim 0.1\mbh$ and radius
$\approx 1.6\mathrm{pc}$. In spite of its larger mass compared to the stellar
disc, the CND is relatively less effective in pushing the stars to extreme
eccentricities due to its larger radius. Smaller efficiency is, on the other
hand, balanced by larger number of stars that are influenced. Hence, another
$\sim100$ stars from the region below $1.5\mathrm{pc}$ may have been tidally
disrupted due to the gravity of the CND.

%uvod, odhady trhani, precese -> rozhazeni disku
%\citep{}, \citet{}, \citealt{}

\section{Conclusions}
Gas in the galactic nuclei can manifest itself not only by means of its
own radiation. It forms complex environment together with the supermassive
black hole and millions of stars. All these components mutually interact and
influence themselves. In this contribution we have focused on the effects
of the gravitational and hydrodynamical influence of the gas in the form
of accretion disc or torus upon the stars. We have discussed the effect of
Kozai oscillations and shown that they are likely to lead to enhanced rate
of tidal stellar disruptions. Due to the relativistic pericentre advance,
this process is strongly damped for $\mbh\gtrsim10^7\msun$, while it is
the most efficient for intermediate masses of the black holes, i.e. it
can help them to grow from the IMBH to the SMBH stage.

Passages of the stars through the gas dissipate their kinetic energy. This
leads to their continuous inflow towards the black hole. The mass flow in
the form of stars has to be small, compared to the accretion rate of the gas,
otherwise, the gaseous structure would be destroyed. The individual stars
may, however, be detectable in the final phase of their inspiral due to
the emission of gravitational waves.

Recent observations of the Galactic Centre indicate recent star formation
at a distance of $10^5 - 10^6\rg$ from the SMBH. This goes in line with
previous theoretical considerations about fragmentation of the outer parts
of the accretion discs due to self-gravity \citep{collin99}. In addition,
recent numerical simulations of the fragmenting accretion flow suggest
that star formation may inhibit further inflow of the gas onto the central
mass \citep{nayakshin07}. Hence, the feedback from stars should be considered
in realistic models of AGNs.

\begin{acknowledgements}
This work was supported by the Research Program MSM0021620860 of the Czech
Ministry of Education and the Centre for Theoretical Astrophysics in Prague.
\end{acknowledgements}

\bibliographystyle{aa}

\end{document}